\documentclass[pdflatex,sn-basic]{sn-jnl}

\usepackage{titlesec}
\usepackage{natbib}
\usepackage{enumitem}
\usepackage{tabularx}
\usepackage{booktabs}
\usepackage{hyperref}
\usepackage{bookmark}
\usepackage{graphicx}%
\usepackage{multirow}%
\usepackage{amsmath,amssymb,amsfonts}%
\usepackage{amsthm}%
\usepackage{mathrsfs}%
\usepackage[title]{appendix}%
\usepackage{xcolor}%
\usepackage{textcomp}%
\usepackage{manyfoot}%
\usepackage{booktabs}%
\usepackage{algorithm}%
\usepackage{algorithmic}
\usepackage{listings}%
\usepackage{bookmark}
\usepackage{anyfontsize}

\renewenvironment{table}[1][]%
{\tableorg[#1]%
\tablebodyfont%
\renewcommand\footnotetext[2][]{{\removelastskip\vskip3pt%
\let\tablebodyfont\tablefootnotefont%
\hskip0pt\if!##1!\else{\smash{$^{##1}$}}\fi##2\par}}%
}{\endtableorg}

\usepackage{float}

\begin{document}

\title[Article Title]{Retrospective Economic Evaluation of Group Testing in the COVID-19 Pandemic}

\author*[1]{\fnm{Michael} \sur{Balzer}}\email{michael.balzer@uni-bielefeld.de}

\author[1]{\fnm{Kainat} \sur{Khowaja}}

\author[1]{\fnm{Christiane} \sur{Fuchs}}

\affil[1]{\orgdiv{Bielefeld University}, \orgname{Faculty of Business Administration and Economics, Data Science Group}, \orgaddress{\street{Universitätsstraße 25}, \city{Bielefeld}, \postcode{33615}, \state{NW}, \country{Germany}}}

\abstract{Surveillance of diseases in a pandemic is an important part of public health policy. Diagnostic testing at the individual level is often infeasible due to resource constraints. To circumvent these constraints, group testing can be applied. The economic cost evaluation from the payer's perspective typically focuses only on deterministic costs which overlooks the substantial economic impact of productivity losses resulting from quarantine and workplace disruptions. The objective of this article is to develop a mathematical model for a retrospective economic evaluation of group testing that incorporates both deterministic costs and income-based economic loss. Group testing algorithms are revisited and simulated at optimized pool sizes to determine the required number of tests. Income data from the German Socio-Economic Panel are integrated into a mathematical model to capture the economic loss. Afterward, hybrid Monte Carlo experiments are conducted by evaluating the economic cost in the Coronavirus disease 2019 pandemic in Germany. Monte Carlo experiments show that the optimal choice of group testing algorithms changes substantially when income-based economic losses are included. Evaluations considering only deterministic costs systematically underestimate the total economic cost. Algorithms with a longer quarantine duration are less attractive than shorter quarantine duration if income-based economic loss is accounted for. The findings show that current evaluations underestimate the true economic cost. Group testing algorithms with shorter duration and fewer stages are preferred, even when they require a larger number of tests. These results underscore the importance of incorporating income-based economic loss into a mathematical model.}


\maketitle

\section{Introduction} \label{title:intro}
A wide range of pathogens cause infectious diseases which have a profound impact on human life. The human immunodeficiency virus, chlamydia, gonorrhea or more recently the late severe acute respiratory syndrome coronavirus 2 (SARS-CoV-2) are examples of such infectious diseases \citep{pilcher2005, datta2007, mutesa2021}. Unsurprisingly, an important part of public health policy consists of the surveillance of diseases where the goal is to monitor and prevent the rapid spread in populations \citep{kline1989}. 

Monitoring leverages diagnostic testing where the most common approach relies on testing individuals in a population by specifically designed protocols. For instance, in pandemics, testing focuses mainly on ill individuals, whereas asymptomatic or healthy individuals are neglected. However, if the population to be screened is large, the individual testing procedure is often difficult or even impossible to perform given limited resources \citep{millioni202}. Alternatively, individuals of a population can be grouped and evaluated on joint test results. If a test indicates that there is no infected individual in a group, then every individual in that group is directly declared as negative and thus non-infected. Consequently, should a group include at least one infected individual, then this group is labeled as positive and further testing is necessary. This algorithmic outline is generally called group testing (GT) \citep{dorfman1943, hwang1972, aldridge2022b}. 

Since the advent of the COVID-19 pandemic, GT has garnered increased attention due to the decrease in test numbers compared to individual testing. Additionally, GT algorithms have been associated with a reduction in economic cost and the mitigation of resource constraint \citep{eberhardt2020, mutesa2021, aldridge2022b}.
For instance, several studies examining SARS-CoV-2 testing in Germany conclude that the economic cost savings can be traced back to a reduction in material costs which are composed of start-up, pre-screening questionnaire and apps as well as implementation costs (see, for example, \cite{nguyen2021, simas2021, nguyen2023, pighi2023}). 

On the negative side, GT algorithms are accompanied by waiting times for testing rounds where individuals from positive pools have to be quarantined until declared negative. However, the economic loss associated with the quarantine duration is unknown. Although \cite{brandt2024} account for the economic loss associated with the quarantine duration in a retrospective healthcare study based on sick benefits, a general economic evaluation of the GT algorithms from the payer's perspective is missing.

The aim of this article is to propose a mathematical model for an economic evaluation of GT algorithms from the payer's perspective in Germany which includes regional, economic and temporal disparities. Furthermore, hybrid Monte Carlo experiments are conducted which combine artificial generation of optimal test numbers for GT algorithms with a real-world income data set for the economic loss. Particularly, the mathematical model relies on interpreting the monetary compensation of individuals in the form of their income as the economic loss for the quarantine duration. Since the incomes in Germany are subject to variation due to spatio-temporal disparities, often unavailable, outdated or subject to measurement error, the data set provided by the Socio-Economic Panel (SOEP) research infrastructure is utilized. The data set consists of a large-scale panel data study on households and individuals via questionnaires \citep{soep2022}.

Motivated by conjectures in GT literature which suggest that GT algorithms with longer quarantine duration are generally unlikely to be applied in pandemics, the hybrid Monte Carlo experiments show that focusing on purely deterministic costs leads to an underestimation of economic costs \citep{aldridge2022b}. Thus, by tackling GT algorithms under the assumption that they are embedded in real-world application settings where resource-constraints and uncertainty are naturally occurring, the results confirm that GT algorithms with shorter quarantine duration are preferred.

In Section \ref{title:methods}, the mathematical model is introduced. Afterwards, Monte Carlo experiments are conducted in Section \ref{title:res}. Finally, results are concluded and discussed in Section \ref{title:dis}.

\section{Methods} \label{title:methods}
\subsection{Group Testing algorithms}
In Section \ref{title:intro}, GT is described as any algorithm that divides a population into disjoint pools and evaluates results based on joint tests. A variety of GT algorithms have been developed which follow this algorithmic outline. The focus in this article lies on well established GT algorithms which garnered profound interest in pandemics. Particularly, multi-stage Dorfman-type algorithms discussed in \citep{dorfman1943} and \citep{patel1962} are considered. The practical application in the COVID-19 pandemic of these GT algorithms has been reviewed in \cite{eberhardt2020, mutesa2021} and \cite{aldridge2022b}. 

Generally, GT algorithms can be divided into two types, which are adaptive and non-adaptive algorithms. Moreover, GT algorithms can be divided further into two different branches, which are probabilistic and combinatorial GT algorithms. Non-adaptive GT algorithms are based on designing the testing procedure beforehand such that intermediate results do not matter in subsequent stages. Combinatorial GT is more general and assumes a deterministic model but is also often not applicable in pandemics due to the inherent complex procedure and long-running time of the algorithms. Adaptive GT algorithms rely on stage-wise procedures, where further tests can only be carried out when the results from the previous testing stages are recovered. Probabilistic GT relies on the assumption that there is a probability distribution on the number of infected individuals \citep{sobel1959,hwang1972,du1999,aldrige2019,mutesa2021}. Multi-stage Dorfman-type algorithms can be described as adaptive, probabilistic GT algorithms, where in the final stage individual retesting has to be performed to confirm the status of infection. Such GT algorithms are called trivial or conservative and can more easily be investigated mathematically \citep{aldridge2022b}.

Generally, computing the number of tests is non-trivial due to the inherent probabilistic nature in multi-stage Dorfman-type algorithms. However, under certain assumptions, the expected number of tests can be derived which is useful to find optimal pool sizes. In terms of notation, let $p \in (0,1)$ be the prevalence, $q = 1-p$ the complement, $n \in \mathbb{N}$ the number of individuals in a population and $s_l$ the size of the pools in stage $l \in \{1,\dots,k\}$ with $k \in \mathbb{N}$ being the maximum number of stages. Furthermore, denote by $T_k$ a random variable with distribution $F_{T_k}$ describing the number of tests. Then $\mathbb{E}(T_k)$ is the expected number of tests. In order to obtain an expression for $\mathbb{E}(T_k)$, assumptions are imposed: The population size $n$ is finite and large, the epidemic evolves dynamically but the prevalence $p$ is constant. The absolute number of infected individuals is given by $p \cdot n$ such that the assumption of the independent and identical probability for the infection of individuals holds. Individuals in the population do not interact and tests are perfectly accurate, that is, $100\%$ sensitivity and specificity holds.

\begin{figure}[!htpb] 
    \centering
    \small
    \includegraphics[scale = 0.5]{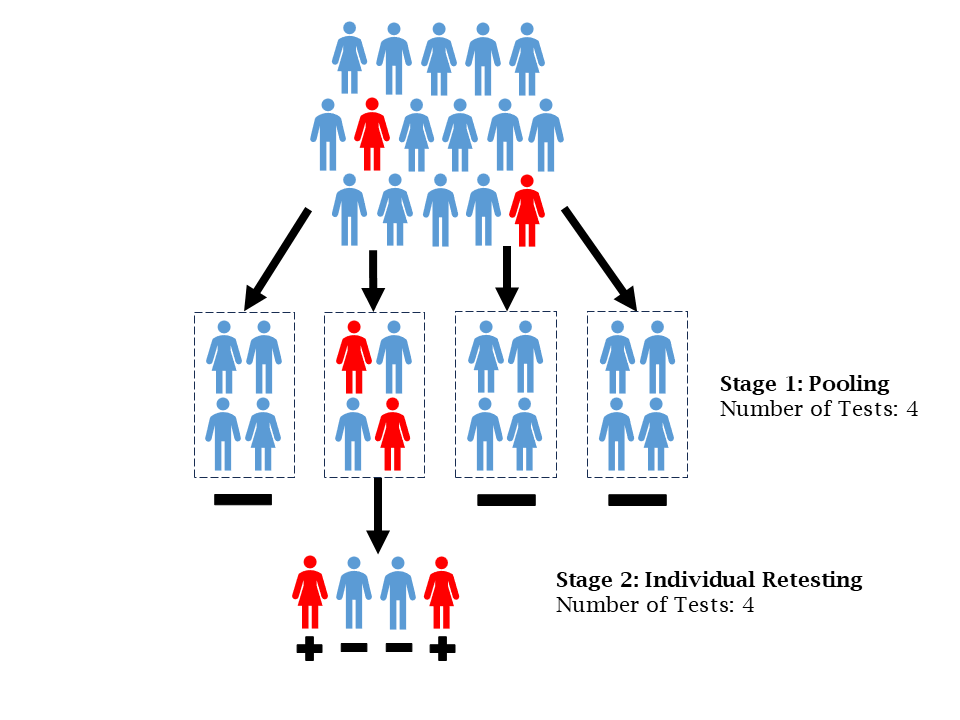} \hfill
    \caption{Illustrative scheme of the two-stage group testing (GT) algorithm according to \cite{dorfman1943}. The considered population size is $n = 16$ with pool sizes $s_1 = 4$. Red indicates the infected and blue the non-infected individuals.}
    \label{fig:dorfman}
\end{figure}

The Dorfman algorithm is a two-stage GT algorithm which starts by combining samples derived from individuals into $\frac{n}{s_1}$ disjoint pools of equal size $s_1$, ignoring the fact that $\frac{n}{s_1}$ is not necessarily a whole number. Afterward, the states of the individuals are evaluated based on these joint tests. Thus, a positive joint test indicates that the corresponding group is positive, meaning that at least one individual in this pool is infected such that additional testing is necessary. Individuals that are part of a negative pool are directly declared as not infected. In the two-stage GT algorithm, follow-up individual tests are conducted in the second stage to confirm the infectious status of the individuals. The expected number of tests is thus given by 
\begin{equation}
    \mathbb{E}(T_2) = n\left(\frac{1}{s_1} + 1 - q^{s_1}\right).
\end{equation}
An illustration of the Dorfman algorithm can be seen in Figure \ref{fig:dorfman} \citep{dorfman1943}.

In multi-stage Dorfman-type algorithms, the second stage of individual retesting is postponed in favor of additional joint test stages. Therefore, samples from individuals in positive pools are again grouped into $\frac{n(1-q^{s_1})}{s_2}$ pools of size $s_2$ and the results are evaluated based on joint tests results. In this process, individuals who are tested in the same pool at stage~$k$ have already been in a joint pool at stage~$k-1$. In the final stage of the multi-stage GT algorithms, individual retesting of the leftover individuals has to be performed. By construction, the test results always depend on the previous testing outcomes. The expected number of tests in the three-stage GT algorithm is given by 
\begin{equation}
        \mathbb{E}(T_3) = n\left(\frac{1}{s_1} + \frac{1}{s_2} (1-q^{s_1}) + (1-q^{s_2})\right).
\end{equation}

In the four-stage GT algorithm, individual retesting is additionally postponed in favor of a joint test stage. Samples from individuals in positive pools are then grouped once again into $\frac{n(1-q^{s_1}) (1-q^{s_2})}{s_3}$ pools of size $s_3$ and the results are evaluated based on joint tests results. The expected number of tests is then given by 
\begin{equation}
        \mathbb{E}(T_4) = n\left(\frac{1}{s_1} + \frac{1}{s_2} (1-q^{s_1}) +  \frac{1}{s_3}(1-q^{s_2}) + (1-q^{s_3})\right).
\end{equation}
Alternatively, an additional joint test stage can be conducted which effectively yields a five-stage GT algorithm. The expected number of tests can be recovered via
\begin{equation}
         \mathbb{E}(T_5) = n\left(\frac{1}{s_1} + \frac{1}{s_2} (1-q^{s_1}) +  \frac{1}{s_3}(1-q^{s_2}) + \frac{1}{s_4}(1-q^{s_3})  + (1-q^{s_4})\right).
\end{equation}
Thus, samples from individuals in positive pools are again grouped into $\frac{n(1-q^{s_1}) (1-q^{s_2})(1-q^{s_3})}{s_4}$ pools, which allows for additional joint tests \citep{patel1962}.

\subsection{Mathematical model for economic cost evaluation}
For the development of a simple mathematical model, the basic microeconomic principle for cost evaluation from a payer's perspective as discussed by \cite{osullivan2003} is followed. In particular, let $\tau_{(p,u)}^{(k)} \in \mathbb{N}$ be the number of tests for a GT algorithm with a maximum number of $k$ stages at prevalence $p$ in geographical location~$u$. Then, the deterministic costs depending on $\tau_{(p,u)}^{(k)}$ from the payer's perspective are given via a linear relationship by 
\begin{equation*}
        c_f + \tau_{(p,u)}^{(k)} \cdot c_v.
\end{equation*}
The initial costs $c_f \in \mathbb{R}^+ $ are independent of the number of tests and can be described as the costs for building the necessary infrastructure and planning for GT algorithms. The variable costs $c_v \in \mathbb{R}^+$ are composed of any arbitrary number of different cost components. For example, these components can be material, personal and logistic costs. The costs are assumed to increase linearly in $\tau_{(p,u)}^{(k)}$ since every additional test results in a fractional increase in the variable cost. The deterministic cost structure is based on empirical case studies in the COVID-19 pandemic as discussed by \cite{simas2021, nguyen2023} and \cite{pighi2023}.

Additionally, the presence of resource constraints occurring in real-world GT applications is included. For instance, these constraints arise in situations where not all tests can be conducted due to limited testing capacities in laboratories or qualified personnel for sample collection. To ensure that all tests can be conducted, an additional cost component is introduced which captures the outsourcing of the leftover tests. Mathematically, an upper bound $\tau_0 \in \mathbb{N}$  for the testing capacity is specified as
\begin{equation*}
\max\!\left(0, \, (\tau_{(p,u)}^{(k)} - \tau_0)\,c_l \right).
\end{equation*}
Exceeding the testing capacity leads to the outsourcing of the leftover tests at higher laboratory costs $c_l \in \mathbb{R}$.

Since the considered GT algorithms have multiple stages and test results are not available instantaneously, individuals have to be quarantined because the status of their infection remains unclear. From the payer's perspective, individuals may be hindered in pursuing their work given their occupation. Therefore, quantifying the economic loss associated with the quarantine duration becomes a central obstacle. In the model mathematical model, the economic loss is defined as the compensation an individual receives despite being unable to work. This definition is motivated by two perspectives. First, since individuals are unable to work, the companies are effectively losing productivity for the duration of the quarantine. This leads to effective adverse effects on the overall production and profit over time. Second, in the COVID-19 pandemic in Germany, the German Infection Protection Act stated that individuals had to be compensated directly from the employer for the quarantine duration. In turn, the employer could claim the paid compensation from the authority \citep{ifsg2024}. 

Two recent healthcare studies utilize a similar definition of economic loss. The assumption of economic loss based on sick pay has been taken into account in an estimation of average COVID-19-associated costs per capita based on empirical health insurance claims for Germany \citep{brandt2024}. Another study specializing on the cost evaluation of testing procedures in the COVID-19 pandemic in emergency rooms in German hospitals takes a similar notion of economic loss into account \citep{diel2022}. 

To mathematically formalize the economic loss and embed it into the mathematical model, the structural hierarchy and spatial variations of incomes have to be considered. For instance, the shape of the income distribution in schools will severely differ from the distribution in large companies. Furthermore, industries are not uniformly distributed in a geographical location. Thus, evaluating the economic cost of GT algorithms in different geographical locations for different economic activities requires detailed information about the incomes. Additionally, the duration of a stage in a GT algorithm has to be quantified. For simplicity, the duration is set to exactly one day in the mathematical model. By construction, every individual in a population has to be tested in the first stage such that all individuals stay in quarantine for one day. In any subsequent stage, only individuals in positive pools remain in quarantine. Let $w_{(j,u)} \in \mathbb{R}^+$ denote the sum of the daily incomes of all individuals in quarantine in any stage $j \in \{1,\dots, k\}$ in geographical location~$u$. Adjusting the economic loss for the proportion of the individuals who can work remotely defined by~$h \in [0,1]$ and combining the economic loss with the deterministic and outsource costs yields the overall economic cost $c_{(p,u)}^{(k)} \in \mathbb{R}^+$ for a GT algorithm as
\begin{equation*}
    c_{(p,u)}^{(k)} = \begin{cases}
          c_f + \tau_{(p,u)}^{(k)} \cdot c_v + (1-h) \cdot \sum_{j = 1}^{k} w_{(j,u)} & \text{if} \
 \tau_0 \geq \tau_{(p,u)}^{(k)} \\
      c_f + \tau_0 \cdot c_v + (1-h) \cdot  \sum_{j = 1}^{k} w_{(j,u)}  +  (\tau_{(p,u)}^{(k)} - \tau_0) \cdot c_l & \text{if} \ \tau_0 < \tau_{(p,u)}^{(k)}.
      \end{cases}
\end{equation*}

\section{Results} \label{title:res}
\subsection{Empirical setting and data collection}
To investigate the effects of changes in parameters in the proposed mathematical model, hybrid Monte Carlo experiments are conducted which combine variables based on real-world data such prevalence and income levels and costs that are predetermined. The empirical setting is concerned with the retrospective economic evaluation of GT in the COVID-19 pandemic in Germany. Three districts, namely Hamburg, Bremen and Berlin-Mitte are chosen to exemplify and demonstrate the effects of variable changes in the mathematical model. The choice of districts is motivated by the available data on prevalence levels as well as incomes. Particularly, information about prevalence levels in the German districts are of primary interest which correspond to the geographical locations in the mathematical model. Thus, a large-scale data set containing seven-day incidence rates for COVID-19 on district level from the Robert Koch Institute is utilized \citep{robertkochinstitut2024}.

\begin{figure}[!htpb] 
    \centering
    \includegraphics[width = \textwidth]{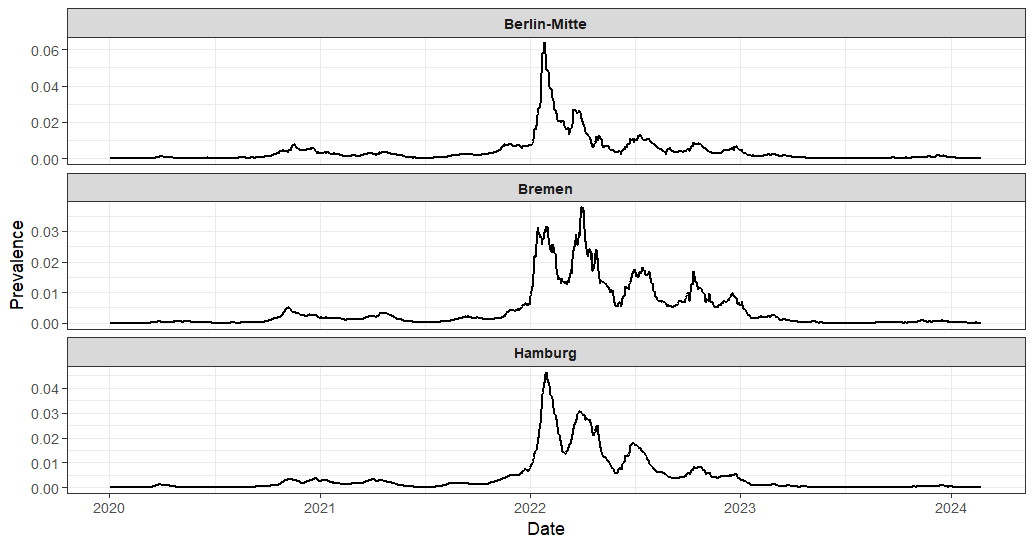} \hfill
    \caption{Progress of approximated point prevalence values in the COVID-19 pandemic in German districts Berlin-Mitte, Bremen and Hamburg across the years 2020 to 2024.}
    \label{fig:prevalence}
\end{figure}

Since prevalence levels are not readily available in the data set, an approximation of the point prevalence has to be computed. The point prevalence can be approximated as the product of the incidence rate and the duration of the infectious disease \citep{bruce2018}. Assuming constant arrival of new cases over a period of seven days combined with a suitable average duration of SARS-CoV-2, the prevalence levels in German districts can be approximated. However, there is no consensus on the average duration of SARS-CoV-2. \cite{cevik2021} provide a systematic review and meta-analysis for the duration of viral shedding in SARS-CoV-2 and find average results ranging from 14 to 126 days. In a different study, \cite{walsh2020} report that COVID-19 patients with mild-to-moderate illness are usually infectious until day ten with some individuals having a longer time frame of infectiousness. Similarly, \cite{rhee2021} observed that patients with mild to moderate illness remain infectious for up to 10 days. For those who are severely to critically ill, the infectious period extends to 15 days, with a maximum duration of 20 days. These results are also supported by the World Health Organization \citep{who2024}. Therefore, 14 days is taken as a reference value in the Monte Carlo experiments. The approximated point prevalence for the districts of interest can be seen in Figure \ref{fig:prevalence}.

\begin{figure}[!htpb] 
    \centering
    \includegraphics[width = \textwidth]{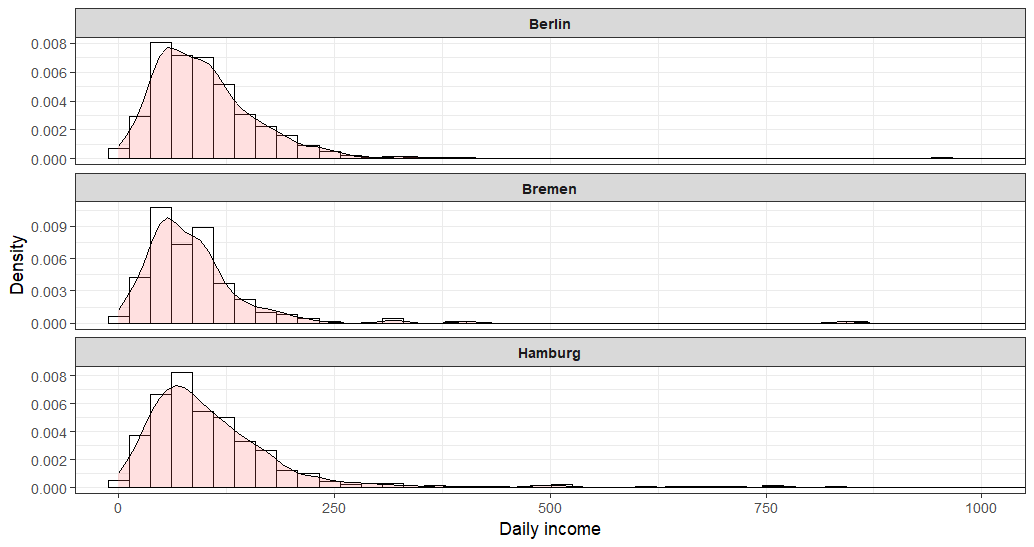} \hfill
    \caption{Histogram and kernel density estimates of daily incomes in EUR (capped at 1000) in Berlin-Mitte, Bremen, and Hamburg in the years 2019 to 2020.}
    \label{fig:incomes}
\end{figure}

The income data set of the SOEP-Core in the most recent version from the year 2021 is additionally utilized. This data set is based on repetitive and representative surveys with questionnaires on households and individuals from the years 1984 to 2021 in Germany. The central variables of interest are the state of residence, the monthly gross labor income and the worked hours in a week of an individual in the years 2019 to 2021 which partially correspond to the ongoing COVID-19 pandemic in Germany. Since exact information about the daily incomes is not readily available in the data set, estimation is necessary. Utilizing the worked hours in a week under the assumption of a typical five day work week on average yields the actual worked daily hours. Analogous, weekly incomes can be obtained by dividing the monthly gross labor income by the average work weeks in a year of $4.345$. By dividing the weekly incomes by the actual daily worked hours, daily incomes are obtained. Thus, the final data set contains $23727$ observations and four variables. The computed daily incomes for the districts of interest can be seen in Figure \ref{fig:incomes}.

\subsection{Study Design}
To stochastically simulate the GT algorithms, pool sizes in different stages have to be chosen. If pool sizes are specified freely, the number of tests can be suboptimal. To ensure that the economic cost are evaluated under optimal number of tests, the expression of the expected number of tests is minimized with respect to unknown pool sizes. Given the size of the pools in optimum, the GT algorithm are stochastically simulated for every prevalence $p$ that is available in the geographical location $u$ across the entire time period and the resulting number of tests $\tau_{(p,u)}^{(k)}$ are computed. Similarly, daily incomes are drawn with replacement from the empirical distribution at random. By construction, daily incomes for subsequent stages of GT are always drawn based on the incomes of quarantined individuals of the previous stage. In total, $n_{\text{sim}} = 25$ simulations are conducted at each prevalence $p$ in which the number of tests $\tau_{(p,u)}^{(k)}$ and economic cost $c_{(p,u)}^{(k)}$ are computed separately.

For the evaluation of the results, a local point-based sensitivity analysis is conducted. In principle, such an analysis is a one-at-a-time technique in which the effect of one changing parameter is evaluated while the other parameter are fixed \citep{razavi2015}. In every case, the base-line model is included for direct comparability between different scenarios which is specified as initial costs $c_f = 10000$, variable costs $c_v = 150$, outsourcing costs $c_l = 300$, population size $n = 1000$, percentage of remote work $h = 0.5$ and testing capacity $\tau_0 = 750$. In Table \ref{tab:val}, parameter values for all Monte Carlo experiments can be found.

\begin{table}[!htbp]
\caption{\label{tab:val} Parameter values in the Monte Carlo experiments for retrospective economic cost evaluation in the COVID-19 pandemic in Germany.}
\centering
\begin{tabular}{cc}
\toprule
\textbf{Parameter} & \textbf{Values} \\
\midrule
$c_f$ &  10000    \\
$c_v$ &  \{0, 50, 100, 150, 300, 400, 800, 1000, 2000\}   \\
$c_l$ &  300     \\
$n$ &  \{150, 250, 500, 1000, 5000, 10000\}  \\
$h$ &  \{0,0.4,0.5,0.8,0.9,1\}  \\
$\tau_0$ &  \{0, 50, 100, 250, 750, 1000\} \\
\bottomrule
\end{tabular}
\end{table}

For the evaluation of the Monte Carlo experiments, performance criteria have to be considered. Particularly, the economic cost per individual (ECI) are defined as 
\begin{equation*}
    \frac{\bar{c}_{(p,u)}^{(k)}}{n}
\end{equation*} 
where $\bar{c}_{(p,u)}^{(k)}$ are the average economic cost at prevalence $p$ in geographical location $u$ across all number of simulations $n_{\text{sim}}$. The ECI can be interpreted as the proportion of average economic cost distributed equally across all individuals in the population. Lower values are always preferred. To quantify the spread of the ECI, the range is additionally provided defined as 
\begin{equation*}
    \frac{R_{(p,u)}^{(k)}}{n} = \frac{c_{(p,u)_{\text{max}}}^{(k)}}{n} - \frac{c_{(p,u)_{\text{min}}}^{(k)}}{n}
\end{equation*}
where $\frac{c_{(p,u)_{\text{max}}}^{(k)}}{n}$ are the maximum and $\frac{c_{(p,u)_{\text{min}}}^{(k)}}{n}$ are the minimum economic cost per individual across simulations.

Monte Carlo experiments are conducted in the programming language R \citep{R}. Particularly, for the optimization of the GT algorithms the \textbf{optimx} package is utilized \citep{nash2011, nash2014}. Data pre-processing and visualization is performed via the \textbf{tidyverse} packages \citep{wickham2019}. For parallel computing, the packages \textbf{furrr} and \textbf{doFuture} in combination with \textbf{foreach} are utilized \citep{bengtsson2021, vaughan2022}. For reproducibility of all results, the code for the functions and simulations can be found in GitHub repository \url{https://github.com/micbalz/EconEvalGT}.

\subsection{Impact of incomes in German districts}
\begin{figure}[!htpb] 
    \centering
    \includegraphics[width = \textwidth]{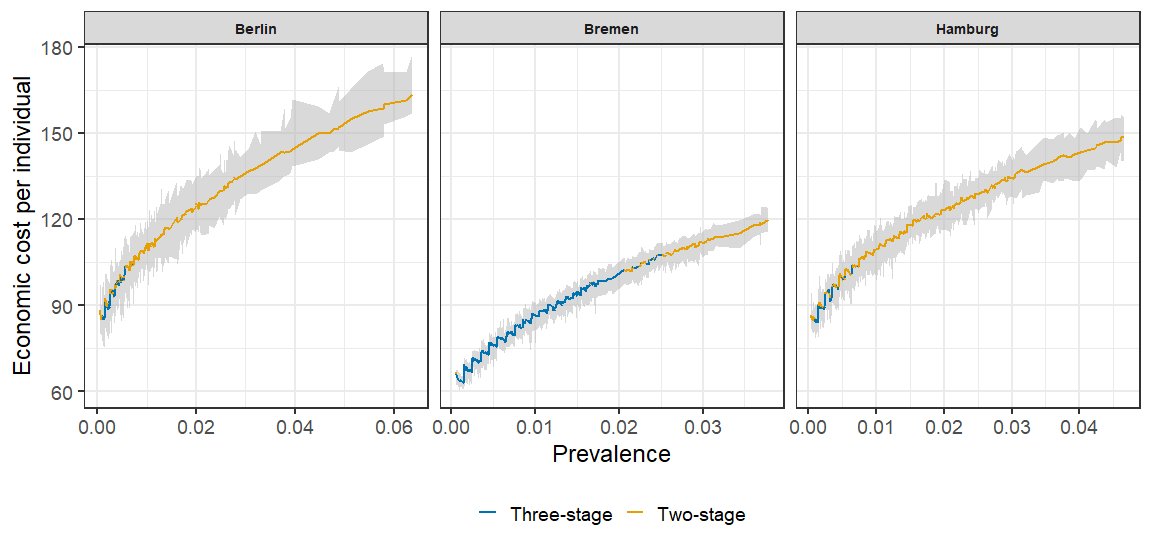} \hfill
    \caption{Progress of economic cost in EUR per individual (ECI) for the COVID-19 pandemic for the districts Hamburg, Bremen and Berlin. Solid lines represents the average ECI. Uncertainty is visualized by the range via shaded areas around the average ECI.}
    \label{fig:ctcost}
\end{figure}

In Figure \ref{fig:ctcost}, the effects of different income levels on the ECI in German districts are presented. The results for Berlin indicate that the ECI ranges from $75$ to $155$ over different prevalence values. The three-stage algorithm has the lowest average ECI until prevalence values of $0.005$. However, due to the randomness when drawing incomes from the empirical distribution, the choice shifts occasionally towards the two-stage algorithm. For prevalence values larger than $0.005$, the preferred choice gradually shifts towards the two-stage algorithm.

In contrast, the results for Bremen show that the ECI is generally lower, ranging from values $60$ to $120$. The lowest average ECI can thus be observed in the three-stage algorithm until prevalence values of $0.03$. Afterwards, the preferred choice gradually favors the two-stage algorithm. However, the results for Hamburg are closely aligned to the results of Berlin. The largest deviation can be seen in the range where Berlin exhibits larger uncertainty than Hamburg. Similar results for Berlin and Hamburg are unsurprising since the distributions of the incomes in Figure \ref{fig:incomes} show that the mean incomes in both districts are similar. However, the deviation in the range can be attributed to fact that the standard deviation of incomes in Berlin is higher which leads to an increase in the range for Berlin.

In general, the results indicate that higher incomes shift the preference in terms of GT algorithms towards fewer numbers of stages. Thus, lower incomes indicate that the adverse effect of the additional quarantine duration in economic cost is weaker. Combined with lower number of tests, the results show the preference towards three-stage algorithms. Similarly, the range is much narrower due to the lower incomes. Thus, depending on the incomes in a district, payers should utilize two- and three-stage algorithms for lowering their economic cost. If incomes are high, the two-stage algorithm is preferred in order to clear individuals faster. Payers with lower incomes can perform GT for more numbers of stages but should not choose a GT algorithm with more than three stages.

\subsection{Impact of population size}
\begin{figure}[!htpb] 
    \centering
    \includegraphics[width = \textwidth]{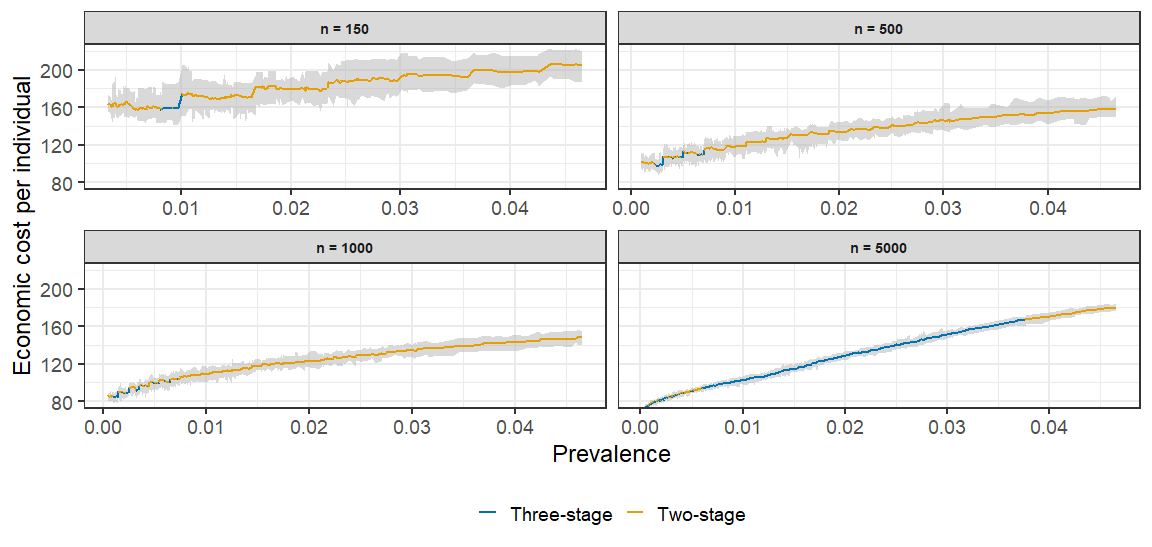} \hfill
    \caption{Progress of economic cost in EUR per individual (ECI) for the COVID-19 pandemic in Hamburg for changing population size $n \in \{150,500,1000,5000\}$. Solid lines represents the average ECI. Uncertainty is visualized by the range via shaded areas around the average ECI.}
    \label{fig:popcosts}
\end{figure}

In Figure \ref{fig:popcosts}, the results for the impact of varying population sizes can be seen. Given the similarity of the prevalence values in the considered districts, the focus lies on Hamburg for brevity in the presentation of the results.

For a population size of $n = 150$, the ECI ranges from $140$ to $220$ throughout the prevalence values. Almost always, the two-stage algorithm is the optimal choice. Notable is the erratic behavior occurring in the average ECI. The erratic behavior can be explained by the underlying permutations in GT algorithms if population sizes are small. For instance, depending on the optimal pool size, an additional infected individual found in a separate, otherwise negative pool leads to a doubling in the number of tests if all the other infected individuals are condensed in one pool. Consequently, a larger number of individuals has to be quarantined for a prolonged time. Since incomes are drawn randomly, an individual with a higher wage leads to a non-smooth behavior in the average ECI. Additionally, the range is large. The optimal choice of the two-stage algorithm remains for population sizes of $n = 500$ and $n = 1000$. However, the most notable difference is the decrease in range as well as the ECI due to the weakening of the marginal income effect. For $n = 5000$, the optimal choice changes towards the three-stage algorithm almost entirely whereas the ECI has increased in comparison to a population size of $n = 1000$. Furthermore, the erratic behavior has vanished entirely and the average ECI curve is smooth.
With an increasing population size, the effect of economic loss is outweighed by deterministic and outsource costs. Since $\tau_0$ is set to $750$ in the baseline model, the increase in the ECI can be explained by an outsourcing of the majority of tests. Thus, GT algorithms with a smaller number of tests are generally preferred. Consequently, if population sizes are large and testing capacities rigid, then GT algorithms with a longer quarantine duration and lower number of tests are the better option.

\subsection{Changes in deterministic costs}
\begin{figure}[!htpb] 
    \centering
    \includegraphics[width = \textwidth]{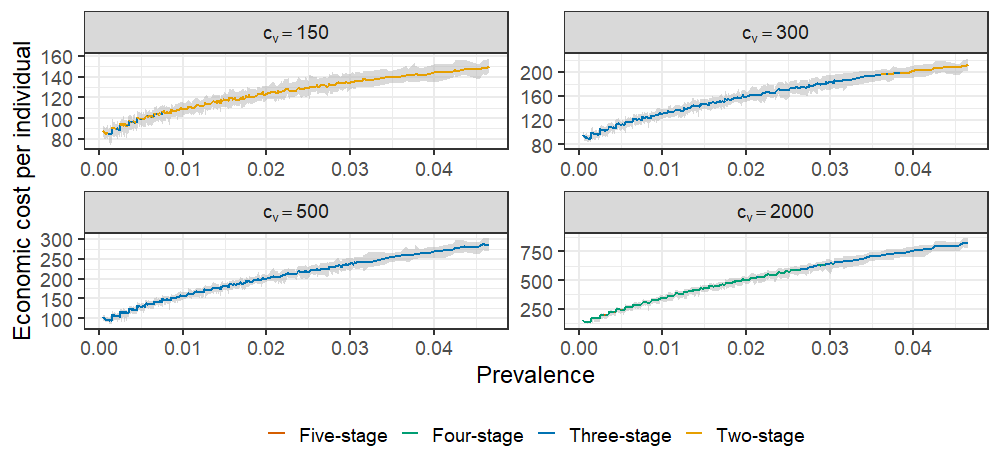} \hfill
    \caption{Progress of economic cost in EUR per individual (ECI) over the COVID-19 pandemic in Hamburg for changing values of variable costs $c_v \in \{150,300,500,2000\}$. Solid lines represents the average ECI. Uncertainty is visualized by the range via shaded areas around the average ECI.}
    \label{fig:detcosts}
\end{figure}

In the following, changes in deterministic costs on the ECI are considered. In principle, changes in the deterministic costs act as a scaling factor in the number of tests. The mathematical model suggests that the optimal choice tends to GT algorithm with lower number of tests as the deterministic costs increase. These expectations are confirmed in Figure \ref{fig:detcosts}.
For $c_v = 150$, the ECI ranges from $80$ to $180$. The two-stage algorithm is almost always preferred which is as expected since it has the lowest quarantine duration and thus the lowest economic loss. Note that individual testing is not the optimal choice due to testing capacity of $\tau_0 = 750$ and population size $n = 1000$ in the baseline model.

As $c_v$ increases, the ECI reaches values of $200$ where the three-stage algorithm is almost always the optimal choice. In this case, the lower number of tests in the three-stage algorithm outweighs the economic loss from the increased quarantine duration. For $c_v = 2000$, the optimal choice changes further to GT algorithms with a larger number of stages. The ECI ranged from $100$ to $750$. The range increases additionally with deterministic costs in absolute values. Overall, the results show that the higher the deterministic costs, the stronger the optimal choice changes towards GT algorithms with a larger number of stages and lower number of tests.

\subsection{Changes in outsource costs}
\begin{figure}[H] 
    \centering
    \includegraphics[width = \textwidth]{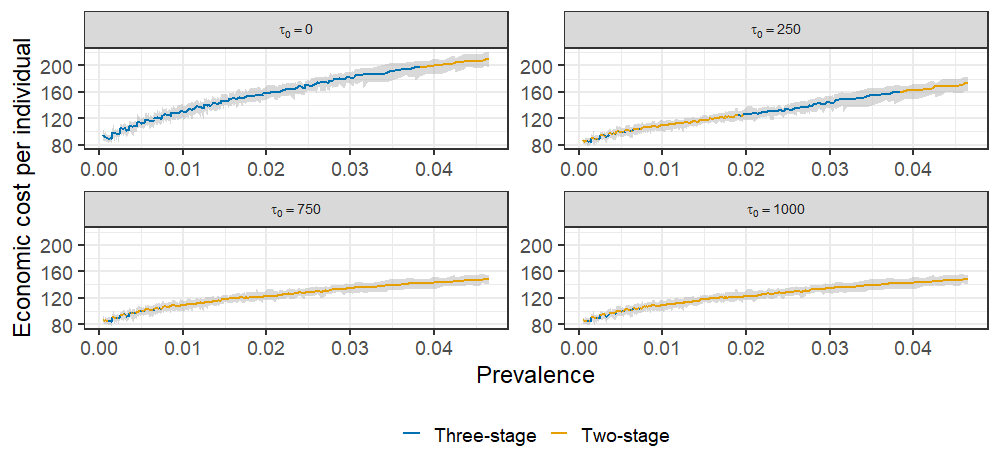} \hfill
    \caption{Progress of economic cost in EUR per individual (ECI) over the COVID-19 pandemic in Hamburg for changing values of the testing capacity $\tau_0 \in \{0,250,750,1000\}$. Solid lines represents the average ECI. Uncertainty is visualized by the range via shaded areas around the average ECI.}
    \label{fig:capcosts}
\end{figure}

Similar to the deterministic costs, changes in outsource costs are investigated in the following. The mathematical model suggests that the testing capacity is similar to the deterministic costs, scaling with the difference between the number of tests and the testing capacity. In principle, GT algorithms with a lower number of tests are preferred as testing capacity decreases. In Figure \ref{fig:capcosts} the expectations can be confirmed.
If $\tau_0 = 0$, that is, there is no internal testing capacity, then every test has to be outsourced. The ECI varies between $80$ and $215$. In this case, the three-stage algorithm is the preferred GT algorithm. Generally, the range is also quite pronounced. Changes on the optimal choice of the GT algorithm can be seen in a testing capacity of $\tau_0 = 250$. 
The results indicate that the number of outsourced tests is low or almost non-existent for multi-stage GT algorithms. However, the two-stage algorithm becomes occasionally the optimal choice. The ECI decreases further to the maximum values of $175$. As $\tau_0 > 250$, the ECI ranges from $80$ to $150$. The optimal choice is almost always the two-stage algorithm. Outsourcing for multi-stage GT algorithms is never conducted since the testing capacity is enough to satisfy the demand for tests. Thus, the two-stage algorithm is preferred due to the higher number of tests and lower quarantine duration. 

\subsection{Effects of Changes in Remote Work}
\begin{figure}[!htpb] 
    \centering
    \includegraphics[width = \textwidth]{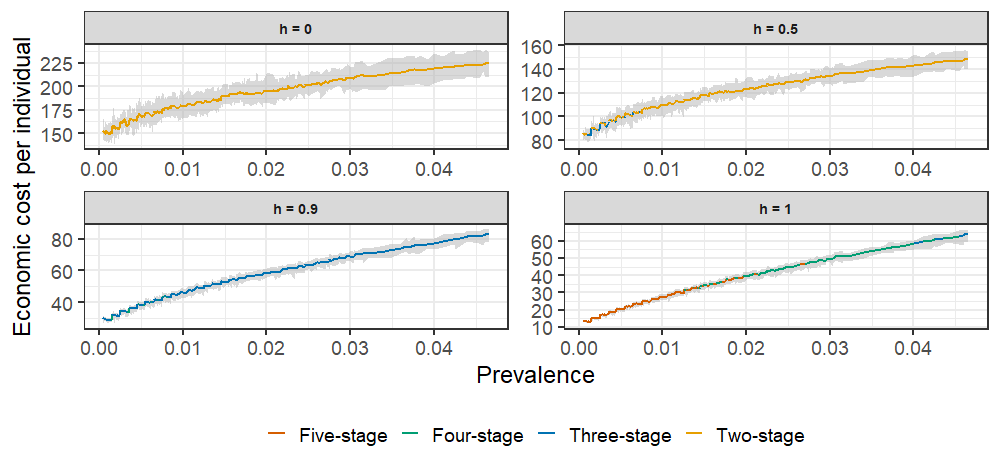} \hfill
    \caption{Progress of economic cost in EUR per individual (ECI) over the COVID-19 pandemic in Hamburg for changing values of the remote work $h \in \{0,0.5,0.9,1\}$. Solid lines represents the average ECI. Uncertainty is visualized by the range via shaded areas around the average ECI.}
    \label{fig:hcosts}
\end{figure}

From the mathematical model, it can be inferred that changes in remote work have an effect on the economic loss. In particular, $h$ can be seen as a scaling factor leading to an increase in the ECI. The expectation can be seen in Figure \ref{fig:hcosts}. 
For $h = 0$, that is, there is no possibility of remote work, the ECI ranges from $140$ to $250$. The optimal choice is always the two-stage algorithm. Thus, the GT algorithm with the lowest quarantine duration is preferred, albeit permitting disadvantages due to larger number of tests. Differences in terms of optimal choice can be seen for $h = 0.5$ where in low to moderate prevalence situations the three-stage algorithm becomes the preferred choice. The strong decrease in economic cost based on the economic loss is the main driver of this change. The number of tests is lower in the three-stage algorithm which paired with the strong decrease in economic loss leads to a change in the optimal choice. For $h = 0.9$ the three-stage algorithm is always the optimal choice. The reasoning remains as before.

The most interesting changes can be seen for $h = 1$. This corresponds to scenarios where every individual in a screened population can work remotely. Consequently, the economic loss is zero. Therefore, this corresponds purely to the evaluation of the deterministic costs. It can be concluded that in low to moderate prevalence situation, the five-stage algorithm is preferred. In moderate to moderately high prevalence situations, the four-stage algorithm is the optimal choice. In the very high prevalence situations, the three-stage algorithm is the optimal choice. Additionally, lowest ECI values are reached ranging from $10$ to $60$. Furthermore, the range has decreased substantially. Unsurprisingly, the best practice from the payer's perspective is to incorporate a high percentage of remote work, if possible. This leads to a longer quarantine duration but since economic loss is non-existent, the lowest ECI due to the lowest number of tests can be ensured. GT algorithms with many stages are the preferred option. If high percentages cannot be ensured, the combinations of two-stage and three-stage algorithms should be the preferred choice because these provide a decent trade-off between number of tests and economic loss.

\section{Discussion}\label{title:dis}
The key findings and main contributions of this study are: (a) Deterministic costs are combined with economic loss for economic evaluation of GT algorithms in pandemics. To this end, a mathematical model of economic cost is introduced. By proposing a general framework for economic loss based on incomes, the inclusion of additional costs that have been reported in empirical studies by \cite{diel2022} and \cite{brandt2024} are accounted for. (b) The income-based economic loss is motivated from two perspective. Quarantined individuals are unable to work. This results in productivity losses from the payer's perspective. Moreover, the compensation can be regulated by law as in the case of Germany with the Infection Protection Act. Thus, economic loss is represented by an increase in the health care spending. (c) In the case study of economic cost in the COVID-19 pandemic in Germany, spatio-temporal variations are accounted for by utilizing a large-scale income data set on state of residence and district level. Additionally, COVID-19 data from the Robert Koch Institute is utilized \citep{soep2022,robertkochinstitut2024}. (d) Hybrid Monte Carlo experiments confirm that focusing solely on deterministic costs in the evaluation of the ECI for different GT algorithms leads to severe underestimation of underlying, not directly observable costs. The results for the baseline model in the Monte Carlo experiments show that the optimal choice of GT algorithms with the lowest ECI consists of a combination of the two- and three-stage algorithm depending on the prevalence levels and districts. Changes in different parameters lead to a change in the optimal choice of GT algorithms. Particularly, for very large populations, the three-stage algorithm whereas for much smaller population the two-stage algorithm is the optimal choice. Changes in deterministic costs change the optimal choice from the two- to the three-stage algorithm. For extreme values of deterministic costs, the optimal choice changes further to the four- and the five-stage algorithm. Furthermore, these changes are accompanied by an increase in the ECI. Changes in the testing capacity have the exact reverse effect. With an increasing testing capacity, the favored GT algorithm shifts from three- to two-stage. Moreover, this change is accompanied with a decrease in the ECI. The most interesting case can be observed when the remote work parameter $h$ is varied. In this case, the results indicate that high values of remote work lead to a shift in favor of algorithms with a higher number of stages. As the percentage of remote work decreases, the decision changes to GT algorithms with three or two stages. Individual testing is the optimal choice only if the prevalence values are high or very high. Since the overall prevalence in the scenarios are always low, individual testing as the optimal choice has not been observed. 

In the course of the study, several limitations have become apparent. First, strong assumptions in Section \ref{title:methods} that need to hold in order to derive the expressions for the expected number of tests are imposed. Particularly, the assumptions of perfect testing conditions is not realistic but quite convenient for the derivation of the expected number of tests in GT algorithms. A natural extension is to relax some assumptions and investigate how the results change. Second, the GT literature recently received an explosion in work with the development of novel GT algorithms. It is clear that a departure from established GT algorithms considered in this article is another direction for future research. Particularly, the so called grid and hypercube algorithms are promising candidates \citep{mutesa2021, aldridge2022b}. Third, the perspective of economic loss and thus an increased burden for the health care sector relies on regulated compensation by law. For countries that do not introduce a law that ensures the compensation of individuals due to quarantine, only the productivity loss can be taken into account. The focus on Germany is primarily motivated by the compensation of individuals in quarantine which is written in law \citep{ifsg2024}. Results for different countries are also of interest. On the same note, the case study exemplifies mainly the districts of Berlin, Bremen and Hamburg. Different results are obtained if the district and thus prevalence as well as income levels change. Fourth, a simplified form of epidemic dynamics is considered while deliberately ignoring any possible interventions as, for example, vaccines, recovered individuals or asymptomatic infected individuals. Furthermore, a simple, linear mathematical model is constructed. However, non-linear relationships in the model might describe real-world outcomes better. Fifth, only a one-way effect of epidemics has been considered on the ECI. It is also possible to evaluate scenarios where epidemics and the economy interact in non-trivial ways. For instance, if the testing capacity at one day is not reached, the leftover tests can be utilized in later GT algorithms. As such, the economic situation has an effect on the epidemic state where companies have to generate a base revenue to perform GT algorithms. Therefore, these feedback relationship should be considered in a more sophisticated mathematical model. Sixth, besides economic cost, the effect of a prolonged quarantine duration on the mental health can be explored. For example, the mental state of individuals might have an effect on their performance in their occupation. These questions are left open for further research.

\backmatter

\bmhead{Supplementary information}
The code for reproducibility of all Monte Carlo experiments is freely available at the GitHub repository \url{https://github.com/micbalz/EconEvalGT}.

\section*{Declarations}

\bmhead{Funding}
Financial support by the German Research Foundation (DFG) [RTG 2865/1 – 492988838] is gratefully acknowledged.

\bmhead{Data availability}
The COVID-19 data sets supporting the conclusion of this article is available in the Robert-Koch-Institute GitHub repository at \url{https://github.com/robert-koch-institut/COVID-19_7-Tage-Inzidenz_in_Deutschland} (accessed on November 14th, 2023). The German Socio-Economic Panel dataset supporting the conclusion of this article can be requested from the German Socio-Economic Panel research infrastructure at \url{https://www.diw.de/en/diw_01.c.615551.en/research_infrastructure__socio-economic_panel__soep.html}. 

\bmhead{Conflict of interests}
The authors declare that they have no conflict of interest.

\bibliography{sn-bibliography}

@article{kline1989,
	Title = {Evaluation of human immunodeficiency virus seroprevalence in population surveys using pooled sera},
	Author = {Kline, RL and Brothers, TA and Brookmeyer, R and Zeger, S and Quinn, TC},
	DOI = {10.1128/jcm.27.7.1449-1452.1989},
	Number = {7},
	Volume = {27},
	Month = {July},
	Year = {1989},
	Journal = {Journal of clinical microbiology},
	ISSN = {0095-1137},
	Pages = {1449-1452}
}

@article{pilcher2005,
author = {Pilcher, Christopher D. and Fiscus, Susan A. and Nguyen, Trang Q. and Foust, Evelyn and Wolf, Leslie and Williams, Del and Ashby, Rhonda and O'Dowd, Judy Owen and McPherson, J. Todd and Stalzer, Brandt and Hightow, Lisa and Miller, William C. and Eron, Joseph J. and Cohen, Myron S. and Leone, Peter A.},
title = {Detection of Acute Infections during HIV Testing in North Carolina},
journal = {New England Journal of Medicine},
volume = {352},
number = {18},
pages = {1873-1883},
year = {2005},
doi = {10.1056/NEJMoa042291},
note ={PMID: 15872202}
}

@article{millioni202,
  title={Test Groups, Not Individuals: A Review of the Pooling Approaches for SARS-CoV-2 Diagnosis},
  author={Millioni, Renato and Mortarino, Cinzia},
  journal={Diagnostics},
  volume={11},
  year={2021},
  pages={68},
  doi={10.3390/diagnostics11010068}
}

@article{dorfman1943,
 ISSN = {00034851},
 author = {Robert Dorfman},
 journal = {The Annals of Mathematical Statistics},
 number = {4},
 pages = {436--440},
 publisher = {Institute of Mathematical Statistics},
 title = {The Detection of Defective Members of Large Populations},
 urldate = {2023-11-15},
 volume = {14},
 year = {1943}
}

@article{mutesa2021,
  title={A pooled testing strategy for identifying SARS-CoV-2 at low prevalence},
  author={Mutesa, L. and Ndishimye, P. and Butera, Y. and Souopgui, J. and Uwineza, A. and Rutayisire, R. and Ndoricimpaye, E. L. and Musoni, E. and Rujeni, N. and Nyatanyi, T. and Ntagwabira, E. and Semakula, M. and Musanabaganwa, C. and Nyamwasa, D. and Ndashimye, M. and Ujeneza, E. and Mwikarago, I. E. and Muvunyi, C. M. and Mazarati, J. B. and Nsanzimana, S. and Ndifon, W.},
  journal={Nature},
  volume={589},
  number={7841},
  pages={276--280},
  year={2021},
  doi={10.1038/s41586-020-2885-5},
}

@article{nguyen2021,
title = {Optimal pooled testing design for prevalence estimation under resource constraints},
journal = {Omega},
volume = {105},
pages = {102504},
year = {2021},
issn = {0305-0483},
doi = {https://doi.org/10.1016/j.omega.2021.102504},
author = {Ngoc T. Nguyen and Ebru K. Bish and Douglas R. Bish}
}

@article{simas2021,
  title={Pooling for SARS-CoV2 Surveillance: Validation and Strategy for Implementation in K-12 Schools},
  author={Simas, Allyson M. and Crott, Josh W. and Sedore, Christine and Rohrbach, Alison and Monaco, Amy P. and Gabriel, Stacey B. and Lennon, Niall and Blumenstiel, Brendan and Genco, Caroline A.},
  journal={Frontiers in Public Health},
  volume={9},
  pages={789402},
  year={2021},
  doi={10.3389/fpubh.2021.789402},
}

@article{nguyen2023,
  title={Cost and cost-effectiveness of four different SARS-CoV-2 active surveillance strategies: evidence from a randomised control trial in Germany},
  author={Nguyen, H. T. and Denkinger, C. M. and Brenner, S. and Koeppel, L. and Brugnara, L. and Burk, R. and Knop, M. and B{\"a}rnighausen, T. and Deckert, A. and De Allegri, M.},
  journal={The European journal of health economics : HEPAC : health economics in prevention and care},
  volume={24},
  number={9},
  pages={1545--1559},
  year={2023},
  doi={10.1007/s10198-022-01561-8},
}

@article{pighi2023,
  title={Cost-effectiveness analysis of different COVID-19 screening strategies based on rapid or laboratory-based SARS-CoV-2 antigen testing},
  author={Pighi, Laura and Henry, Brandon M. and Mattiuzzi, Camilla and De Nitto, Sara and Salvagno, Gian Luca and Lippi, Giuseppe},
  journal={Clinical Chemistry and Laboratory Medicine},
  volume={61},
  number={9},
  pages={e168--e171},
  year={2023},
  doi={10.1515/cclm-2023-0164},
}

@book{aldrige2019,
  author={Aldridge, Matthew and Johnson, Oliver and Scarlett, Jonathan},
  title={Group Testing: An Information Theory Perspective},
  year={2019},
  publisher = {Now Foundations and Trends},
  doi={10.1561/0100000099}}

@inbook{aldridge2022b,
author="Aldridge, Matthew
and Ellis, David",
title="Pooled Testing and Its Applications in the COVID-19 Pandemic",
chapter="Pooled Testing and Its Applications in the COVID-19 Pandemic",
bookTitle="Pandemics: Insurance and Social Protection",
year="2022",
publisher="Springer International Publishing",
address="Cham",
pages="217--249",
isbn="978-3-030-78334-1",
doi="10.1007/978-3-030-78334-1_11",
}

@article{hwang1972,
  title={A method for detecting all defective members in a population by group testing},
  author={Hwang, Frank K.},
  journal={Journal of the American Statistical Association},
  volume={67},
  number={339},
  pages={605--608},
  year={1972},
  doi={10.2307/2284447},
}

@article{patel1962,
 ISSN = {00401706},
 author = {M. S. Patel},
 journal = {Technometrics},
 number = {2},
 pages = {209--217},
 publisher = {[Taylor & Francis, Ltd., American Statistical Association, American Society for Quality]},
 title = {Group-Screening with More Than Two Stages},
 urldate = {2023-12-25},
 volume = {4},
 year = {1962}
}

@article{R,
    title = {R: A Language and Environment for Statistical Computing},
    author = {{R Core Team}},
    journal = {{R Core Team}},
    organization = {R Foundation for Statistical Computing},
    address = {Vienna, Austria},
    year = {2024}
  }

@article{datta2007,
  title={Gonorrhea and chlamydia in the United States among persons 14 to 39 years of age, 1999 to 2002},
  author={Datta, S. D. and Sternberg, M. and Johnson, R. E. and Berman, S. and Papp, J. R. and McQuillan, G. and Weinstock, H.},
  journal={Annals of internal medicine},
  volume={147},
  number={2},
  pages={89--96},
  year={2007},
  doi={10.7326/0003-4819-147-2-200707170-00007}
}

@article{soep2022,
  author       = {{German Institute for Economic Research (DIW Berlin)}},
  title        = {{Socio-Economic Panel (SOEP), data for years 1984-2021, SOEP-Core v38, EU Edition}},
  year         = {2022},
  journal    = {{German Institute for Economic Research (DIW Berlin)}},
  doi          = {10.5684/soep.core.v38.1o}
}

@article{robertkochinstitut2024,
  author = {Robert Koch-Institut},
  title        = {7-Tage-Inzidenz der COVID-19-Fälle in Deutschland},
  year         = {2024},
  journal    = {Zenodo},
  address      = {Berlin},
  doi          = {10.5281/zenodo.10862914}
}

@book{du1999,
  title={Combinatorial Group Testing and Its Applications},
  author={Du, Ding-zhu and Hwang, Frank Kwang-ming},
  year={2000},
  publisher={World Scientific Publishing Co Pte Ltd},
  edition={2nd Revised},
  isbn={9789810241070},
  series={Series On Applied Mathematics},
  format={Hardback},
  language={English},
}

@article{sobel1959,
author = {Sobel, Milton and Groll, Phyllis A.},
title = {Group Testing To Eliminate Efficiently All Defectives in a Binomial Sample},
journal = {Bell System Technical Journal},
volume = {38},
number = {5},
pages = {1179-1252},
doi = {https://doi.org/10.1002/j.1538-7305.1959.tb03914.x},
year = {1959}
}

@book{bruce2018,
  title={Quantitative methods for health research: a practical interactive guide to epidemiology and statistics},
  author={Bruce, Nigel and Pope, Daniel and Stanistreet, Debbi},
  year={2018},
  publisher={John Wiley \& Sons}
}

@article{cevik2021,
  title={SARS-CoV-2, SARS-CoV, and MERS-CoV viral load dynamics, duration of viral shedding, and infectiousness: a systematic review and meta-analysis},
  author={Cevik, Muge and Tate, Matthew and Lloyd, Owen and Maraolo, Alberto E and Schafers, Johannes and Ho, Antonia},
  journal={The Lancet. Microbe},
  volume={2},
  number={1},
  pages={e13--e22},
  year={2021},
  doi={10.1016/S2666-5247(20)30172-5},
  publisher={Elsevier}
}

@article{walsh2020,
  title={The duration of infectiousness of individuals infected with SARS-CoV-2},
  author={Walsh, Kieran A and Spillane, Susan and Comber, Larry and Cardwell, Karen and Harrington, Patricia and Connell, Jeff and Teljeur, Conor and Broderick, Niamh and de Gascun, Cillian F and Smith, Susan M and Ryan, Mary and O'Neill, Marie},
  journal={The Journal of infection},
  volume={81},
  number={6},
  pages={847--856},
  year={2020},
  doi={10.1016/j.jinf.2020.10.009},
  publisher={Elsevier}
}

@article{rhee2021,
  title={Duration of Severe Acute Respiratory Syndrome Coronavirus 2 (SARS-CoV-2) Infectivity: When Is It Safe to Discontinue Isolation?},
  author={Rhee, Chanu and Kanjilal, Sanjat and Baker, Michael and Klompas, Michael},
  journal={Clinical Infectious Diseases},
  volume={72},
  number={8},
  pages={1467--1474},
  year={2021},
  month={Apr},
  day={26},
  doi={10.1093/cid/ciaa1249},
  PMID={33029620},
  PMCID={PMC7499497},
  publisher={Oxford University Press}
}

@book{osullivan2003,
  title={Economics: Principles in Action},
  author={O'Sullivan, Arthur and Sheffrin, Steven M.},
  year={2003},
  publisher={Prentice Hall},
  address={Needham, Mass.}
}

@article{ifsg2024,
    author = {{Bundesministerium für Justiz}},
    journal = {{Bundesministerium für Justiz}},
    title = {{Gesetz zur Verhütung und Bekämpfung von Infektionskrankheiten beim Menschen (Infektionsschutzgesetz - IfSG)}},
    year={2021},
    note = {Accessed on April 24, 2024}
}

@article{brandt2024,
  author = {Brandt, F. and Simone, G. and Loth, J. and Schilling, D.},
  title = {COVID-19-associated costs and mortality in Germany: an incidence-based analysis from a payer's perspective},
  journal = {BMC Health Services Research},
  volume = {24},
  number = {1},
  pages = {321},
  year = {2024},
  doi = {10.1186/s12913-024-10838-y}
}

@article{diel2022,
title = {Point-of-care COVID-19 antigen testing in German emergency rooms – a cost-benefit analysis},
journal = {Pulmonology},
volume = {28},
number = {3},
pages = {164-172},
year = {2022},
issn = {2531-0437},
doi = {https://doi.org/10.1016/j.pulmoe.2021.06.009},
author = {R. Diel and A. Nienhaus}
}

@article{eberhardt2020,
title = {Multi-Stage Group Testing Improves Efficiency of Large-Scale COVID-19 Screening},
journal = {Journal of Clinical Virology},
volume = {128},
pages = {104382},
year = {2020},
issn = {1386-6532},
doi = {https://doi.org/10.1016/j.jcv.2020.104382},
author = {J.N. Eberhardt and N.P. Breuckmann and C.S. Eberhardt}
}

@article{who2024,
  author       = {{World Health Organization}},
  journal       = {{World Health Organization}},
  title        = {Coronavirus Disease (COVID-19)},
  year         = 2024,
  howpublished = {\url{https://www.who.int/news-room/fact-sheets/detail/coronavirus-disease-(covid-19)}},
  note         = {Accessed: 2024-06-13}
}

@article{razavi2015,
author = {Razavi, Saman and Gupta, Hoshin V.},
title = {What do we mean by sensitivity analysis? The need for comprehensive characterization of “global” sensitivity in Earth and Environmental systems models},
journal = {Water Resources Research},
volume = {51},
number = {5},
pages = {3070-3092},
doi = {https://doi.org/10.1002/2014WR016527},
year = {2015}
}

@article{nash2011,
    title = {Unifying Optimization Algorithms to Aid Software System
      Users: {optimx} for {R}},
    author = {John C. Nash and Ravi Varadhan},
    journal = {Journal of Statistical Software},
    year = {2011},
    volume = {43},
    number = {9},
    pages = {1--14},
    doi = {10.18637/jss.v043.i09},
  }

@article{nash2014,
    title = {On Best Practice Optimization Methods in {R}},
    author = {{John C. Nash}},
    journal = {Journal of Statistical Software},
    year = {2014},
    volume = {60},
    number = {2},
    pages = {1--14},
    doi = {10.18637/jss.v060.i02},
  }

@article{wickham2019,
doi = {10.21105/joss.01686}, 
year = {2019}, 
publisher = {The Open Journal},
volume = {4},
number = {43}, 
pages = {1686},
author = {Hadley Wickham and Mara Averick and Jennifer Bryan and Winston Chang and Lucy D'Agostino McGowan and Romain François and Garrett Grolemund and Alex Hayes and Lionel Henry and Jim Hester and Max Kuhn and Thomas Lin Pedersen and Evan Miller and Stephan Milton Bache and Kirill Müller and Jeroen Ooms and David Robinson and Dana Paige Seidel and Vitalie Spinu and Kohske Takahashi and Davis Vaughan and Claus Wilke and Kara Woo and Hiroaki Yutani},
title = {Welcome to the Tidyverse},
journal = {Journal of Open Source Software}
}

@article{vaughan2022,
  title = {furrr: Apply Mapping Functions in Parallel using Futures},
  author = {Davis Vaughan and Matt Dancho},
  journal = {Davis Vaughan and Matt Dancho},
  year = {2022},
  note = {https://github.com/DavisVaughan/furrr, https://furrr.futureverse.org/},
}

@article{bengtsson2021,
  author = {Henrik Bengtsson},
  title = {A Unifying Framework for Parallel and Distributed Processing in R using Futures},
  year = {2021},
  journal = {The R Journal},
  doi = {10.32614/RJ-2021-048},
  pages = {208--227},
  volume = {13},
  number = {2},
}

\end{document}